\documentclass[11pt,twoside]{article}
\usepackage{./asp2014}

\aspSuppressVolSlug
\resetcounters

\begin{document}

\title{An interactive, comparative and quantitative 3D visualization system for large-scale spectral-cube surveys using CAVE2}
\author{D. Vohl,$^{1}$
C. J. Fluke,$^{1,2}$
A. H. Hassan,$^{1}$
D. G. Barnes$^{2}$
\affil{$^{1}$Centre for Astrophysics \& Supercomputing, Swinburne University of Technology, 1 Alfred Street, Hawthorn 3122, Australia}
\affil{$^{2}$Monash e-Research Centre, Monash University, 14 Alliance Lane, Clayton 3168, Australia}}

\paperauthor{Sample~Dany Vohl}{dvohl@astro.swin.edu.au}{0000-0003-1779-4532}{Swinburne University of Technology}{Centre for Astrophysics \& Supercomputing}{Hawthorn}{Victoria}{3122}{Australia}
\paperauthor{Sample~Christopher J. Fluke}{cfluke@astro.swin.edu.au}{}{Swinburne University of Technology}{Centre for Astrophysics \& Supercomputing}{Hawthorn}{Victoria}{3122}{Australia}
\paperauthor{Sample~Amr H. Hassan}{ahassan@astro.swin.edu.au}{}{Swinburne University of Technology}{Centre for Astrophysics \& Supercomputing}{Hawthorn}{Victoria}{3122}{Australia}
\paperauthor{Sample~David G. Barnes}{Author3Email@email.edu}{ORCID_Or_Blank}{Monash University}{Monash e-Research Centre}{Clayton}{Victoria}{3168}{Australia}

\begin{abstract}
As the quantity and resolution of spectral-cubes from optical/infrared and radio surveys increase, desktop-based visualization and analysis solutions must adapt and evolve. Novel immersive 3D environments such as the CAVE2 at Monash University can overcome personal computer's visualization limitations. CAVE2 is part advanced 2D/3D visualization space (80 stereo-capable screens providing a total of 84 million pixels) and part supercomputer ($\sim100$ TFLOPS of integrated GPU-based processing power). We present a novel visualization system enabling simultaneous 3D comparative visualization of $\sim100$ spectral-cubes. With CAVE2 augmented by our newly developed web-based controller interface, astronomers can easily organise spectral-cubes on the different display panels, apply real-time transforms to one or many spectral cubes, and request quantitative information about the displayed data. We also discuss how such a solution can help accelerate the discovery rate in varied research scenarios.
\end{abstract}

\section{Large-scale spectral-cube surveys and the knowledge discovery process}
Current and upcoming large-scale extragalactic surveys aim to study galaxy kinematics, formation and environmental effects by capturing large numbers of spectral-cubes\footnote{A spectral-cube is composed of two spatial dimensions along with a spectral or a velocity dimension.}, each containing large numbers of sources. While a large part of the analysis of these surveys will be done using automated pipelines, the human experience remains a key element of the scientific discovery process. To fully comprehend the data, visualization will play a major role. Data visualization is a fundamental, technologically-led process that enables knowledge discovery \citep{HassanFluke2011PASA}. 

\begin{figure}[!htb]
 \centering
\includegraphics[width=12.00cm]{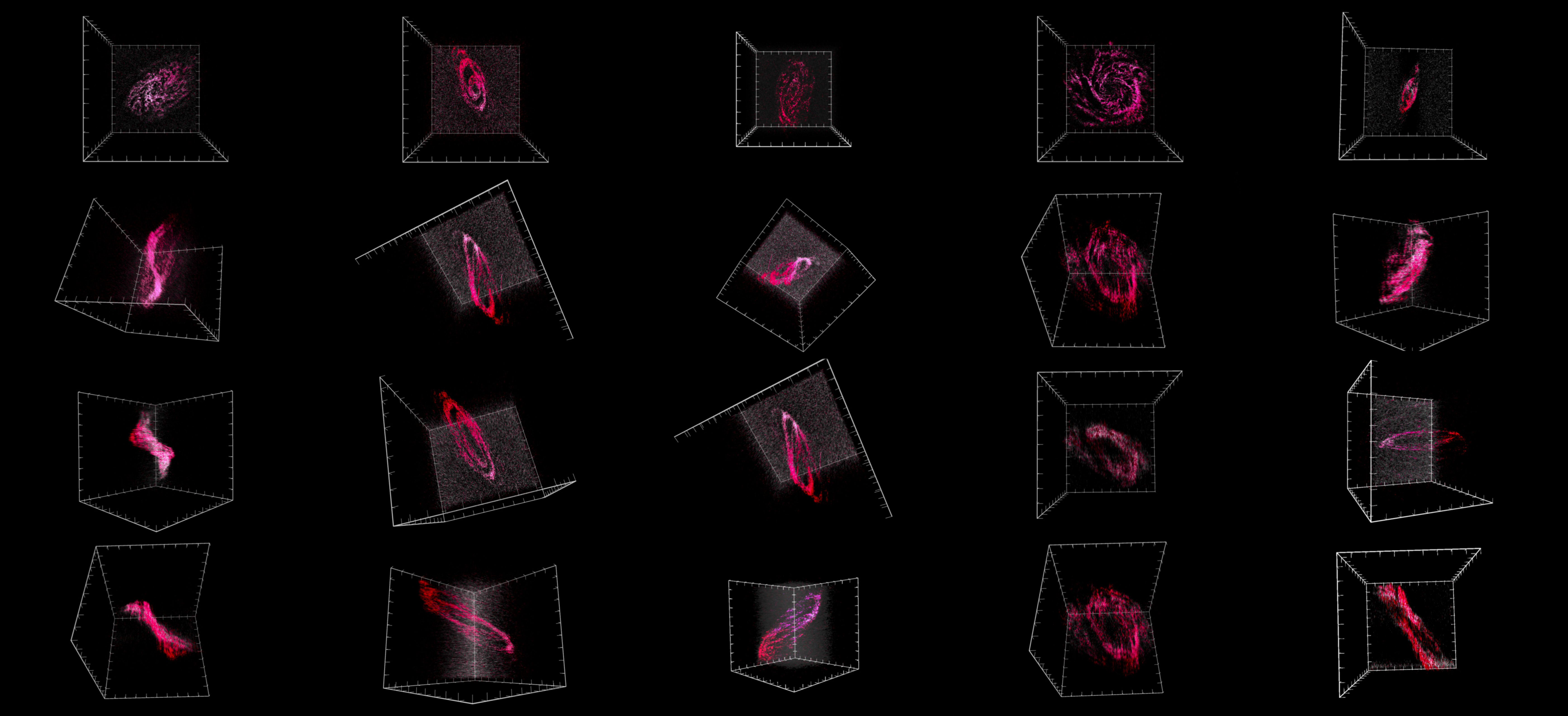}
\caption{Five columns and four rows of CAVE2's 20 columns displaying volume rendered spectral-data from The HI Nearby Galaxy Survey \citep[THINGS;][]{Walter2008AJ}.}
\label{fig::cave2}
\end{figure}

The classical desktop-based visualization methodology where one object is examined at a time is highly inefficient for large-scale spectral-cube surveys. With optical instruments such as SAURON, PMAS/PPAK, SAMI, and MaNGA \citep[e.g. ][]{Sanchez2015IAUS} and radio facilities/instruments such as ALMA, APERTIF, ASKAP, LOFAR, and MeerKAT \citep[e.g. ][]{Johnston2008ExA, Rottgering2011JApA}, data products are growing quickly. As the resolution (voxels or spaxels) of spectral-cubes increases, so too does the number of sources. To give a sense of the size of such surveys, the SAMI Galaxy Survey will observe 3,000 galaxies across a large range of environments, where each spectral-cube is $50^2$ spatial pixels with $2048$ spectral channels \citep{Croom2012MNRAS}. Compare this with the APERTIF radio survey of the northern sky which will record 20,000 spectral-cubes with typical size of $2048^2$ spatial pixels and $16,384$ spectral channels, each containing $\sim100$ sources \citep{Rottgering2011JApA}. 

Instead we look to next generation immersive 3D environments that are designed to deal with large amounts of data in a collaborative setup. The CAVE2 at Monash University is a hybrid 2D/3D reality environment for immersive simulation and information analysis. It is composed of a 8-meter diameter, 320 degree panoramic cylindrical display system using 80 stereo-capable displays arranged in 20 four-panel columns (providing 84 million pixels). With its $\sim100$ TFLOPS of integrated GPU-based processing power, Monash University's CAVE2 is also part supercomputer, making it a great candidate to both visualise a large quantity of data in a collaborative manner and perform compute-intensive data analysis tasks. While such facilities are not yet wide-spread globally, we suspect that they will become more easily available in the future. For large science teams, it may only be necessary for a few team members to access a CAVE2.

\section{An interactive, comparative, and quantitative visualization system}
\begin{figure}[!htb]
 \centering
\includegraphics[width=12.00cm]{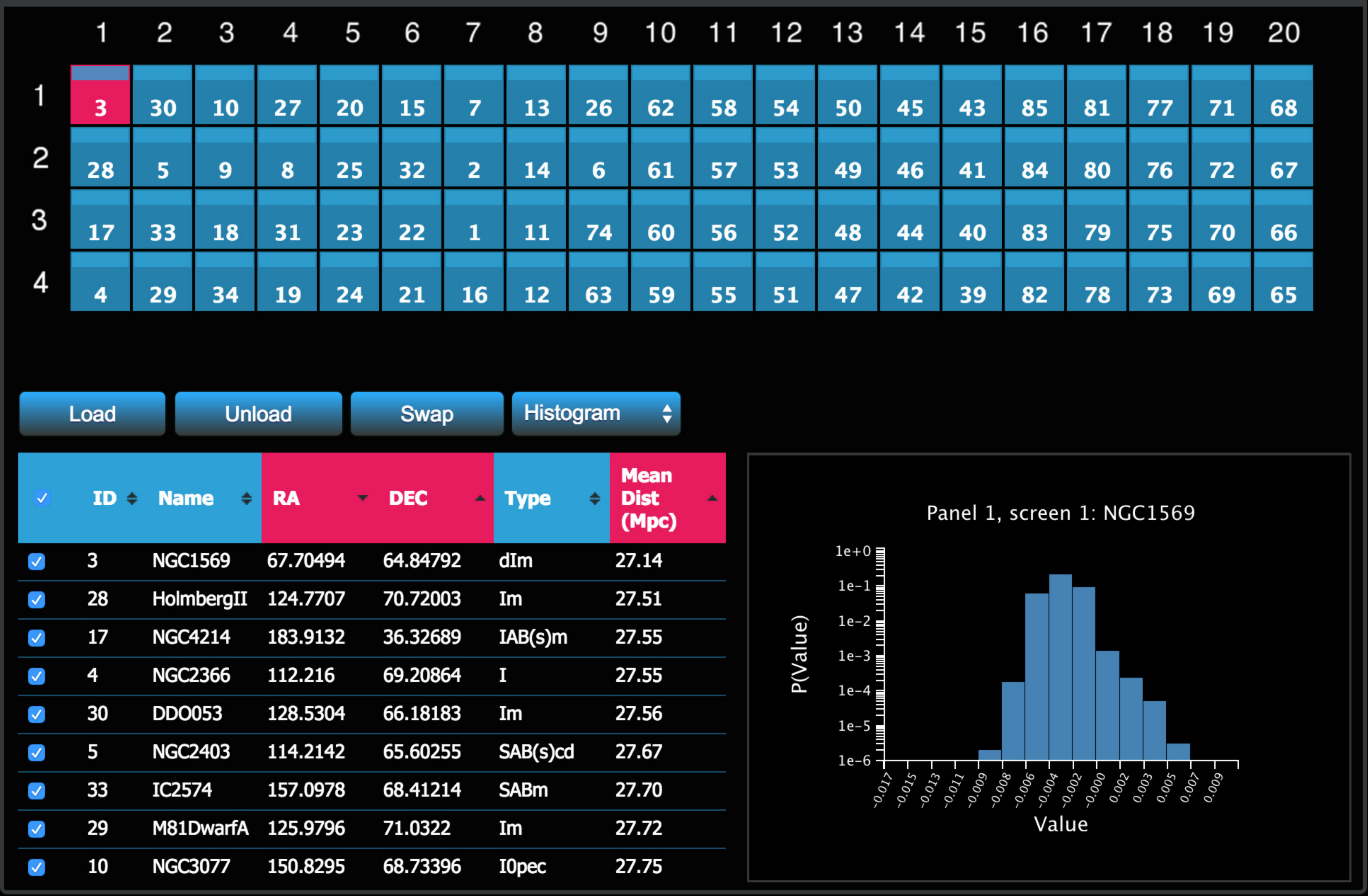}
\caption{A screenshot of the controller: a miniature representation of the CAVE2's 20 four-screen columns state (top); sort data set by multiple criteria (bottom left); and request/display quantitative information (e.g. histogram) about data from one or multiple screens (bottom right).}
\label{fig::controller}
\end{figure}

There is currently no out-of-the-box solution to visualise large catalogues in the CAVE2. To efficiently visualise a large number of spectral-cubes, we developed a visualization system to make the most of the CAVE2's large amount of display area (Figure \ref{fig::cave2}). Each display panel provides a stereoscopic view. Our solution enables the inspection of $\sim100$ spectral-cubes at once at high resolution ($1366 \times 768$ pixels per 3D panel). In such a setting, classical immersive environment controllers such as the wand are not efficient. Hence, we also developed a controller comprising a miniature representation of the CAVE2 (Figure \ref{fig::controller}) which enables a natural interaction between the astronomer, the data, and the visualization environment. Our approach allows quantitative and comparative visualization of many individual spectral data cubes, or multi-wavelength views of a single object. 

The system can be abstracted into three main components: process, render and display (PRD) nodes; a server node; and a web client. 

\emph{PRD nodes}. All processing, rendering, and displaying of volumetric data is done using S2PLOT \citep{BarnesFlukeBourkeParry2006PASA}, an open source library based on OpenGL and written in C. Each PRD node runs one instance of our custom S2PLOT program. This program is responsible for displaying dynamic, interactive 3D geometry including isosurface and texture-based volume rendering. Each PRD node has a specific address to which the server node can connect and send commands to modify the rendered geometry. Given that much of the available processing power lives on the PRD nodes, derived data products or quantitative information such as a moment map or an histogram is computed there.

\emph{Server node}. The server node has three different functionalities. First, it acts as a Content Management System, where a structured list of the data set is available. Secondly, it acts as a Communication Scheduler/Hub, managing communication between the PRD nodes and the web client. Communication with the different PRD nodes is done via TCP sockets using a custom queue, while communication with the web client is done via HTTP methods (request and response). Finally, it acts as a Web Server, enabling the web client to access structured data and the web interface. Throughout a CAVE2 session, the server keeps track of the many states of the different PRD nodes, synchronizing the web page and the PRD nodes. 

\emph{Web client}. A web interface acts as the main controller. A main advantage of using a web interface is that it is light-weight and vastly portable. Multiple users could eventually access the data and control the displays collaboratively, using their own devices such as smart phones, tablets and laptops. 

The controller's functionalities include: 
\begin{itemize}
	\item{interactive miniature CAVE2 allowing a user to}
	\begin{itemize}
		\item{sort a data set by multiple criteria, }
		\item {bulk load/unload data,}
		\item{drag and drop data onto a specific screen,} 
		\item{request derived data products or quantitative information;}
	\end{itemize}
	\item{interactive 3D controller enabling to pan, zoom, or rotate individual or subsets of objects in 3D space;}
	\item{real-time modification of volume's shaders and isosurface parameters}
	\item{slicer with linked views in XY (e.g RA-DEC), XZ (e.g. RA-Velocity) and YZ (e.g. DEC-Velocity).}
\end{itemize}

\section{Accelerating visualization of large-scale surveys}
Our implementation shows that such a solution is feasible in real world settings. Hence, immersive environments such as CAVE2 are prime candidates to help accelerate the discovery rate in the era of large-scale spectral-cube surveys. It is worth noting that such a solution is not only viable for astronomy, but for any volumetric scientific data surveys (e.g. medical imaging). Our solution is also adaptable to other large tiled-display walls. 

We will evaluate the usability of our solution with different science cases, including human-machine assisted morphological classification of galaxies and their environment, accelerated labeling of spectral-cube data for supervised learning, fast evaluation of source finding candidates from large spectral-cubes, and comparative studies of multi-wavelength data.

\acknowledgements 
Thanks to Monash e-Research Centre, Monash University for providing access to the CAVE2. Thanks to ImageHD and CaveHD teams.

\bibliography{P111}  

\end{document}